\documentstyle[prl,aps,12pt]{revtex}
\begin{document}
\draft

\title{Quantitative Determination of a Microscopic Mechanism of High
  Tc Superconductivity}
\author{Eugene Demler and Shou-Cheng Zhang}
\address{
Department of Physics,
Stanford University
Stanford, CA 94305
}
\date{\today}
\maketitle

{\bf 
A major challenge in theoretical attempts to understand the
microscopic mechanism of high $T_c$ superconductivity is to
quantitatively
account for the superconducting condensation 
energy\cite{Schrieffer,Loram1,Loram2,Chester,ILT,Leggett}.
Microscopic model commonly used to describe the superconducting 
copper-oxides, the $t$-$J$ model\cite{ZhangRice}, is beleived to
capture the essential physics of these materials: 
the interplay of electrons kinetic 
energy and their antiferromagnetic exchange interaction.
Within the $t$-$J$ model, the condensation energy can
be
related to the change in the dynamical spin structure factor between
the superconducting and the normal states\cite{ScalapinoWhite}. 
By analyzing previous experimental data, we show that the change 
associated with the resonant neutron scattering 
peak\cite{resonance0,resonance1,resonance2}
found in $YBa_2Cu_3O_7$ can quantitatively account for the
condensation energy. We argue that this analysis 
suggests a microscopic mechanism for high $T_c$
superconductivity, where antiferromagnetic exchange energy is saved
in the superconducting state
through the coupling to a particle-particle
resonance\cite{prl95,prl97-2,Eder,prb98,science}. 
}

One reason for the absence of a generally accepted microscopic
explanation of high $T_c$ superconductivity (HTSC) is the lack of a
quantitatively precise definition of a ``mechanism''. In the
traditional BCS \cite{Schrieffer} superconductors, the phonon
mechanism is quantitatively verified by the isotope substitution 
and single particle tunneling experiments. It is a widely held view
that the HTSC involve strong electronic interaction, while phonon
mediated interaction is not the major driving force for
superconductivity. To date, many different ``mechanisms'' have been
proposed, but very few quantitatively falsifiable predictions have
been checked against experimental data.

A system undergoes a transition from the normal to the 
superconducting
state because it can lower the total free energy. The condensation
energy $E_C$, defined as the energy difference between the normal
state, extrapolated to zero temperature, ($E_N$) and the 
superconducting state ($E_S$), can be directly measured
through the thermodynamical 
critical field $H_c$ by the following relation \cite{Schrieffer}:
\begin{eqnarray}
E_C = E_N - E_S = \frac{H_c^2}{8\pi} V_0
\label{Ecdef}
\end{eqnarray}
Here  we choose $V_0=a\times b\times c$ to be the volume of a unit
cell (in the high $T_c$ cuprates $a=b$ with reasonably good accuracy)
and define the energies inside one unit cell . Alternatively, the
condensation energy can also be found by integrating the difference in
specific heats in the (extrapolated) normal and 
superconducting states from $T=0$ to the
superconducting transition temperature $T_c$ \cite{Loram1,Loram2}.

On the other hand, the condensation energy can be computed in
principle from a microscopic Hamiltonian. Ideally, one would like to
start from the full Hamiltonian involving electronic kinetic energy,
the Coulomb interaction energy and the electron lattice coupling
energy. In fact Chester \cite{Chester}
studied condensation energy of such a Hamiltonian and
showed how various contributions to the condensation energy
could in principle be determined experimentally.
Among many theories of HTSC, only the interlayer tunneling 
mechanism \cite{ILT} made a quantitatively
falsifiable prediction about the condensation energy.  More
recently, Leggett showed that the change in the interlayer Coulomb
energy can be directly related to the frequency and momentum integral
of the dynamic charge susceptibility, a quantity which can be measured
experimentally \cite{Leggett}. He argued that a microscopic theory of
HTSC should 
quantitatively predict the frequency and the momentum range at which
the change of the dynamical susceptibility occurs, and quantitatively
account for the condensation energy.

However, the energy scale involved in HTSC is much smaller than the
Coulomb interaction. In order to understand the mechanism for HTSC, it
is useful to start with an effective Hamiltonian which correctly
describes the basic physics below the Coulomb energy scale. Recently,
Scalapino and White \cite{ScalapinoWhite}
initiated the investigation of condensation energy
in the $t$-$J$ model, which is defined by
\begin{eqnarray}
{\cal H} = -t \sum_{\langle i j \rangle} c_{i\sigma}^{\dagger}
c_{j\sigma} + J \sum_{\langle i j \rangle} {\bf S}_i {\bf S}_j
\label{tJ}
\end{eqnarray}
Here $c_{i\sigma}^{\dagger}$ is electron creation operator on site
$i$ with spin $\sigma$, ${\bf S}_i$ is the electron spin operator, 
${\langle i j \rangle}$ denotes a pair of near neighbor
sites and the Hamiltonian acts on the space of no doubly occupied sites.
Equation (\ref{tJ}) describes the Hamiltonian of one layer only. In a
bi-layer system like $YBa_2Cu_3O_{6+x}$, there are two layers per
unit cell,  
with additional antiferromagnetic (AF) coupling $J_{\perp}$ between the
layers. If a system described by this Hamiltonian undergoes
a 
superconducting transition away from half-filling, the kinetic energy $E_t$
is expected to increase slightly, but the decrease in the exchange
energy $E_J$
may be significantly larger, so that a 
superconducting ground state is realized.

In a tight binding model, the mean kinetic energy $E_t$ can be
expressed as a 
frequency integral of the optical conductivity
$\sigma(\omega)$ \cite{Shastry,SWZ,sudip}. Then the contribution of $E_t$
to the condensation energy can be measured by 
comparing  $\sigma(\omega)$ in the normal and the 
superconducting states. In the low frequency regime, the 
Drude peak in the normal state collapses into a $\delta$
function peak in the 
superconducting state, usually with
conserved weight \cite{IR}. Therefore, the change in the kinetic
energy 
has to be determined from  $\sigma(\omega)$ in the higher
frequency range.

Scalapino and White \cite{ScalapinoWhite} made a insightful
observation that
the change in the $J$ term in equation (\ref{tJ})  can also be directly
expressed as a frequency and momentum integral of the
dynamic spin structure factor $S(q,\omega)$. For bi-layer materials
such as $YBa_2Cu_3O_{6+x}$ it is convenient to separate the even and
odd parts of $S(q,\omega)$ with respect to bi-layer interchange within
a unit cell
\begin{eqnarray}
S(q,\omega)= S^{even}(q,\omega) cos^2(\frac{q_z d_{Cu} }{2})
+  S^{odd}(q,\omega) sin^2(\frac{q_z d_{Cu} }{2})
\label{even-odd}
\end{eqnarray}
Here $d_{Cu}$ is the distance between nearest-neighbor $Cu$-$O$
planes, so $S^{even}$ and $S^{odd}$ describe 
the in-phase and out-of-phase spin fluctuations in the two planes
constituting one bi-layer. Then  $\Delta E_J = E_{J}^{N} - E_{J}^{S} $
may be obtained as ( per unit cell )
\begin{eqnarray}
\Delta E_J = \frac{3}{2} \left( \frac{a}{2 \pi} \right)^2 \int d^2 q
&\int& \frac{d (\hbar \omega)}{\pi}~ \{~\left[ S^{even}_N (q,\omega) - S^{even}_S
  (q,\omega) \right] \left[ \frac{1}{2} J \left( cos (q_x a) + cos (q_y a )
  \right) + \frac{1}{4} J_{\perp} \right]  \nonumber\\
&+& \left[ S^{odd}_N (q,\omega) - S^{odd}_S
  (q,\omega) \right] \left[ \frac{1}{2} J \left( cos (q_x a) + cos (q_y a )
  \right) - \frac{1}{4} J_{\perp} \right]~ \}
\label{DeltaEJ}
\end{eqnarray}
From \cite{Reznik,Hayden} it is known that $J_{\perp}$ is much smaller than $J$,
so the $J_{\perp}$ term in equation (\ref{DeltaEJ}) may be safely
neglected. Scalapino and White
pointed out\cite{ScalapinoWhite} that since
$S(q,\omega)$ can be measured directly in neutron scattering experiments,
$\Delta E_J$ can therefore in principle be measured. If $\Delta E_J$ 
turns out to be much smaller than the experimentally measured 
condensation energy $E_c$, this gives a convincing way to rule out any
mechanism of high $T_c$ superconductivity based on the $t-J$ model.
On the other hand, if $\Delta E_J$ turns out to be greater than 
$E_c$, one would have identified a important driving force for 
superconductivity, and can 
quantitatively test any theoretical mechanism of HTSC based on the
$t$-$J$ model, 
which should predict where in frequency and momentum
space the change of $S(q,\omega)$ occurs. 

However, one should apply this line of reasoning with extreme care. The
quantity $S_N(q,\omega)$ in equation (\ref{DeltaEJ}) is not the normal
state
spin structure factor above $T_c$, but rather an extrapolated normal
state
quantity at $T=0$. Experimentally, one has to carefully identify features
in $S(q,\omega)$ which changes 
abruptly at $T_c$.
Theoretically, one has to identify contributions to $S(q,\omega)$ which
are absent in the normal state and only present in the superconducting state,
so that the 
difference in equation (\ref{DeltaEJ}) can be 
rigorously defined. 
Recently, a resonant neutron peak has been discovered in the family of
$YBa_2Cu_3O_{6+x}$ superconductors \cite{resonance0,resonance1,resonance2}. In the
optimally 
doped $YBa_2Cu_3O_7$ 
superconductor with $T_c=91K$, the resonance peak occurs in the spin
scattering channel at energy $\omega_0=41~meV$. It is centered around
momentum $Q = (\frac{\pi}{a},\frac{\pi}{a})$ and it occurs in the
odd channel with respect to bi-layer interchange. Perhaps the most
remarkable feature of the resonance peak is its onset at the 
superconducting transition temperature $T_c$. Above $T_c$,
there is no experimentally identifiable spectral weight in this
frequency and momentum range. Therefore, this experiment is ideally
suited for analyzing its contribution to the condensation energy.

We constructed a theory \cite{prl95} of the neutron resonance peak by
first identifying a resonance in the spin-triplet particle-particle
(p-p) channel defined by the following operator
\begin{eqnarray}
\pi^{\dagger} = \sum_k ( cos k_x - cos k_y ) c_{k+Q\uparrow}^{\dagger}
c_{-k\uparrow}^{\dagger}  
\end{eqnarray}
In a series of exact diagonalization studies on the Hubbard and
$t$-$J$ models, Meixner, Hanke and the two of us \cite{prl97-2} and
Eder, Hanke and one of us (SCZ) \cite{Eder} demonstrated the existence
of such a triplet p-p resonance for all doping range. A triplet p-p
resonance can not be
detected in the neutron scattering experiments on the normal
state. However, below $T_c$, the d-wave 
superconducting order parameter
mixes a p-p resonance into the dynamical spin structure factor, leading
to a sharply defined collective mode in the triplet particle-hole
(p-h) channel (the so-called $\pi$-resonance). The contribution 
from the p-p resonance to
the p-h channel below $T_c$ is given\cite{prl95,prb98} by
\begin{eqnarray}
\int d( \hbar \omega )~ S(q,\omega)  = \frac{2 | \langle \Delta_d
  \rangle |^2 }{ 1-n } 
\end{eqnarray}
where $\langle \Delta_d \rangle$ is the dimensionless $d$-wave 
superconducting order parameter (not the energy gap) and $n$ is the density of
electrons. This simple estimate of 
the spectral weight of the $\pi$ resonance at $q=Q$ 
agrees reasonably well with
the absolute intensity measured by 
Fong {\it et al.}\cite{resonance2}.
On the other hand, the width of the $\pi$ resonance in
momentum space can not be calculated reliably. Within the $SO(5)$
theory\cite{science}, the $\pi$ resonance has been interpreted as a
pseudo-Goldstone boson of the spontaneous $SO(5)$ symmetry breaking in
the superconducting state. Its existence is therefore directly
correlated with the existence of the superconducting long range order. 

Since the
coupling to a p-p resonance is only possible in the superconducting state, it indeed
contributes to the difference in $S(q,\omega)$ as required by
equation (\ref{DeltaEJ}).
In this work we wish to point out that the logic leading to our
explanation of the $\pi$ resonance below $T_c$ can be reversed to
provide
a microscopic mechanism of HTSC. On one hand, the
existence of the $\pi$ resonance is a unique property of the
superconducting state, on the other hand, its existence around
$Q$ lowers $E_J$, as we see from equation (\ref{DeltaEJ}).
Therefore, the system undergoes a 
superconducting transition, so that a $\pi$  
resonance mode can emerge, which in turn lowers $E_J$.
In other words, no matter how antiferromagnetically correlated the
normal state is, a superconducting state can always further 
lower the exchange energy by coupling a p-p resonance into 
the dynamic spin structure factor $S(q,\omega)$.
Our mechanism for HTSC makes a quantitatively precise prediction which
can be checked experimentally. Once we assume that the HTSC is
dominantly driven by the 
emergence of a new collective mode, the $\pi$ resonance, this
mechanism predicts that the condensation energy is 
determined by the zero temperature spectral weight of the $\pi$
resonance, and the energy saving occurs precisely in the frequency and
momentum range of the $\pi$ resonance.

Since our mechanism predicts an 
additional spectral weight around
$q=Q$ in the superconducting state, a careful reader may wonder what
happens to the spectral sum rule, which states that $S(q,\omega)$
integrated over both $q$ and $\omega$ (without the 
$f(q)=cos (q_x a) + cos(q_y b)$ factor in equation (\ref{DeltaEJ}))
should be a constant, independent of the nature of the ground state.
Of course our additional spectral weight should be compensated by
a depletion of spectral weights from other parts in the momentum space.
However, since $f(q)$ has a absolute minimum at $q=Q$, their
contributions
to (\ref{DeltaEJ}) can not completely cancel the $\pi$ resonance
contribution. 
In fact, depletion of the spectral weight near $q=0$ 
would lead to
additional enhancement of the condensation energy.

Since both the condensation energy and the spectral weight of
the neutron resonance peak have been measured experimentally, it is
straightforward to check this prediction. Within Landau-Ginzburg
theory,
$
H_c = \frac{\Phi_0}{2 \sqrt{2} \pi \xi \lambda}
$,
where $\Phi_0 = \frac{h c}{2 e}$ is the 
superconducting flux quantum,
$\xi$ is the coherence length and $\lambda$ is the London
penetration depth. In $YBa_2Cu_3O_7$, these two length scales are
determined
to be in the range of $\xi =12-20 \AA$ and $\lambda = 1300-1500
\AA$. Using $a=b=3.85~\AA$ and $c=11.63~\AA$ this gives a condensation
energy of $E_C = 3.3-12~ K$ per unit 
cell. On the other hand, $E_C$ can also be measured directly in
specific heat experiments. Loram et al.
reported (see page 251 of \cite{Loram1} and page 247 of \cite{Loram2}) a value
of $E_C= 6~K$ per unit cell for $YBa_2Cu_3O_7$. Let us now turn to the
experimental measurement of  
$\Delta E_J$ given by equation (\ref{DeltaEJ}). In practice, it is
very hard to measure the change of $S(q,\omega)$ throughout the
frequency and momentum range. However, since our mechanism predicts
that the major part of the condensation energy is saved in the range
of $\omega_0 = 41~meV$ and a 2D momentum transfer $ Q =
(\frac{\pi}{a},\frac{\pi}{a})$, we 
can analyze the neutron scattering data by assuming that 
the contributions of the other parts
do not 
change significantly at $T_c$. 
Indeed, for frequencies above the resonance the neutron scattering intensity 
does not change as the system goes from normal to superconducting. For frequencies 
below and in the resonance range the normal state intensity
$S_N(q,\omega)$ is below experimental sensitivity limit
in both even and odd channels. Because of the sum rule 
discussed above we may not assume that it vanishes identically,
so we make an assumption that in this frequency range the spectral weight 
$S_N(q,\omega)$ is 
spread uniformly in momentum 
space and therefore does not contribute to (\ref{DeltaEJ}).
Below $T_c$, a neutron
resonance peak emerges in the odd channel, and $S^{odd}_S(q,\omega)$
has been measured in 
absolute units in the range of the resonance.  The dimensionless
quantity $\int d ( \hbar \omega ) S^{odd}_S(Q,\omega)$ is measured to be
$0.52$
at $T=10~K$ by Fong {\it et al.}\cite{resonance2}
(Theoretical estimate based on references \cite{prl95,prb98} gives
0.32 for this value). 
The resonance has a Lorentzian profile centered at $q=Q$ with a
width $\kappa_{2D} =0.23\AA^{-1}$ \cite{resonance2}, so
the  2D momentum integral can be easily estimated. 
\begin{eqnarray}
\Delta E_J &=& \frac{3}{2}~  \pi~ \left( \frac{a}{2 \pi}
  \kappa_{2D} \right)^2 \times \frac{1}{2} \times \frac{0.52}{\pi} \times 2 \times J
= 0.016 J
\label{DEJres}
\end{eqnarray}
Taking $J = 100~meV$ we obtain $\Delta
E_J = 18~K$.  The actual condensation
energy should be smaller than 
$\Delta E_J$ since the kinetic energy is expected to increase below
the superconducting state. Taking this into account, we see that the spectral
weight of the $\pi$ 
resonance peak can quantitatively account for the condensation
energy measured from $H_c$ and specific heat experiments, in reasonable
agreement with the prediction based on our mechanism.

It is worthwhile to compare our mechanism for HTSC with the spin
fluctuation pairing mechanism
\cite{Scalapino,Pines,Schrieffer2}. All these mechanisms are
based on the AF exchange interaction interaction
$E_J$. Our analysis of the neutron data confirms that a large part of
the condensation energy indeed arises from the AF exchange 
interaction\cite{ScalapinoWhite}. However, our
mechanism differs from spin fluctuation pairing mechanism, both
conceptually and quantitatively. The idea of spin fluctuation pairing
is directly borrowed from the phonon mediated pairing in the
traditional BCS superconductors. It requires sizable
AF spin fluctuations in the normal state, in order to pair
the fermi liquid like quasiparticles. In contrast, in our mechanism, the
AF spin fluctuation does not play an important role in driving
superconductivity, and is neglected
in the normal state of
optimally doped $YBa_2Cu_3O_7$ superconductor. In the 
superconducting state,
the $\pi$ resonance can be viewed as a kind of spin fluctuation in a
particular frequency and momentum range. From the point of view of the
$SO(5)$
theory\cite{science}, the superconducting state is obtained 
from the AF state by a 
rotation, it is therefore
natural to expect it to have more AF correlation than the normal state.
Therefore, the crucial conceptual difference between these two
mechanisms lies in the fact that normal state AF spin fluctuations 
are not 
required in our mechanism.
This conceptual difference has a
direct quantitative consequence. From equation (\ref{DeltaEJ}) we
see that the presence of spin fluctuation spectral weight in the normal
state contributes 
negatively to the condensation energy.
It is not clear why the spin fluctuation model would predict a 
change in the frequency integrated weight of $S(q,\omega)$
near $q=Q$. In fact, most theories of the neutron resonance peak
based on the spin fluctuation models\cite{bulut,morr} assume a preexisting
overdamped spin fluctuation mode in the normal state and only predict
a reduction of damping below $T_c$, with essentially conserved weight.


Our mechanism provides a natural explanation to the doping
dependence of the condensation energy.  We
attribute the condensation energy to the difference in exchange
energies in superconducting and normal states.  
More underdoped materials have considerably more AF
correlations in the nomal state, as known from the neutron 
scattering\cite{Hayden2}.
Although in the superconducting state the weight of 
the resonance is enhanced\cite{Fong}, the difference 
between the normal
and the superconducting exchange energies becomes smaller. 
This is consistent with the results of Loram {\it et al.}
who find a decrease in the condensation energy with decreased 
doping in $Y_{0.8}Ca_{0.2}Ba_2Cu_3O_{6+x}$ \cite{Loram3}. 
Until this point we have been discussing
the relation between the condensation energy and the resonant peaks
in neutron scattering at zero temperature. It would be interesting to see
whether this idea works at finite temperature as well. Without going into
any details we would like to point at a striking similarity between
the temperature dependencies of $H_c$\cite{Loram1} and the 
resonance intensity\cite{Fong,Dai} for the underdoped cuprates.
Not only both quantities scale similarly below $T_c$, but they both have 
tails extending above $T_c$, presumablly 
arising from significant pairing fluctuations
in the pseudogap regime of the cuprates. 
We postpone the  
quantitative analysis of these phenomena to future investigations.

We would like to thank P.~Dai, H.~Fong, S.~Heyden, B.~Keimer, H.~Mook, and
D.~Scalapino for useful discussions. 


\begin{thebibliography}{10}

\bibitem{Schrieffer}
Shrieffer, J.R. Theory of Superconductivity, {\em Addison-Wesley,
MA}, (1964).


\bibitem{Loram1}
Loram, J. {\it et al}. The electronic specific heat of $YBa_2(Cu_{1-x}Zn_x)_3O_7$
from $1.6$ to $300K$.  {\em Physica},  {\bf C 171}  243-256 (1990).  

\bibitem{Loram2}
Loram, J. {\it et al}. Electronic specific heat of $YBa_2Cu_3O_{6+x}$
from $1.8$ to $300K$.  {\em Journal of Superconductivity},  {\bf 7}  243-249 
(1994).   
 
\bibitem{Chester}
Chester, G.V. Difference between normal and superconducting states of a 
metal. {\em Phys. Rev.},  {\bf 103}  1693-1699 (1956). 

\bibitem{ILT} Chakravarty, S. {\it et al}. Interlayer tunneling and gap 
anisotropy in high temperature superconductors. 
{\em Science},  {\bf 261} 337-340 (1993). 

\bibitem{Leggett} 
Leggett, A. Where is the energy saved in cuperate superconductivity?
{\em preprint}.


\bibitem{ZhangRice} 
Zhang, F.C. and Rice, T.M., Effective Hamiltonian for the superconducting
copper-oxides.
{\em Phys. Rev. B}, {\bf 37}, 3759 (1988)

\bibitem{ScalapinoWhite} 
Scalapino, D.J. and White, S., The superconducting condensation energy
and an antiferromagnetic exchange based pairing mechanism. 
{\em preprint}, cond-mat/9805075.

\bibitem{resonance0}
Mook, H. {\it et al}. Polarized neutron determination of the magnetic excitation
in $Y Ba_2 Cu_3 O_7 $. {\em Phys. Rev. Lett.},  
{\bf 70} 3490-3493 (1993). 

\bibitem{resonance1} 
Fong, H.F. {\it et al}. Phonon and magnetic neutron scattering at $41$
$meV$ in $Y Ba_2 Cu_3 O_7 $. {\em Phys. Rev. Lett.}, {\bf 75} 316-319
(1995).

\bibitem{resonance2} 
Fong, H.F. {\it et al}. Polarized and unpolarized neutron scattering study
of the dynamic spin susceptibility in $Y Ba_2 Cu_3 O_7 $.
{\em  Phys. Rev. B},   {\bf 54} 6708-6720 (1996).

\bibitem{prl95} 
Demler, E. and Zhang, S.C. Theory of resonant neutron scattering of
high $T_c$ superconductors. {\em Phys. Rev. Lett.} {\bf 76}
4126-4129 (1995).

\bibitem{prl97-2}
Meixner, S. {\it et al}. Finite size studies on the $SO(5)$ symmetry 
of Hubbard model.
{\em Phys. Rev. Lett.},  {\bf 79} 4902-4905 (1997).

\bibitem{Eder}
Eder, R., Hanke, W., and Zhang, S.C. Numerical evidence for $SO(5)$
symmetry and superspin multiplets in the two dimensional $t-J$ model.
{\em Phys. Rev. B.},  {\bf 57} 13781-13789 (1998).

\bibitem{prb98}
Demler, E., Kohno, H. and Zhang, S.C. $\pi$ excitation of the $t-J$
model.
{\em preprint}, cond-mat/9710139.

\bibitem{science}
Zhang, S.C. A unified theory based on $SO(5)$ symmetry of
superconductivity
and antiferromagnetism.
{\em Science}, {\bf 275} 1089-1096 (1997).


\bibitem{Shastry} Shastry, S. and Sutherland, B. Twisted boundary conditions and
effective mass in Heisenberg-Ising and Hubbard rings. {\em
Phys.Rev.Lett.},  
{\bf 65}  243-246 (1990).

\bibitem{SWZ}
Scalapino, D., White, S. and Zhang, S.C. Superfluid density and the Drude
weigth of the Hubbard model. {\em Phys.Rev.Lett.}, 
{\bf 68} 2830-2833 (1992).

\bibitem{sudip}
Chakravarty, S. Do electrons change their c-axis kinetic energy upon
entering the superconducting state? {\em preprint}, cond-mat/9801025.

\bibitem{IR} Timusk, T. and Statt, B. The pseudogap in high temperature 
superconductors: an experimental survey.  {\em Reports of
Progress in Physics}, to be published.

\bibitem{Reznik}
Reznik, D. {\it et al.} Direct observation of optical magnons in $Y Ba_2 Cu_3 O_{6.2}$. 
{\em Phys. Rev. B},  
{\bf 53} R14741-R14744 (1996)

\bibitem{Hayden}
Hayden, S. {\it et al.} High Frequency spin waves in $Y Ba_2 Cu_3 O_{6.15}$. 
{\em Phys. Rev. B}, {\bf 54} R6905-R6908 (1996)


\bibitem{Scalapino} 
Scalapino, D. The case for $d_{x^2-y^2}$ pairing in the cuprate superconductors.
 {\em Phys. Rep. }, {\bf 250} 329-365 (1995).

\bibitem{Pines}
Pines, D. Understanding high-temperature superconductivity, a progress report.
 {\em Physica B}, {\bf 199-200} 300-309 (1994).

\bibitem{Schrieffer2}
Shrieffer, J.R. Ward's identity and the suppression of spin fluctuation superconductivity.
 {\em J.Low Temp. Physics}, {\bf 99} 397-402 (1995).

\bibitem{bulut} Bulut, N. and Scalapino, D. Neutron scattering from a collective spin fluctuation
mode in $Cu O_2$ bilayer. {\em Phys. Rev. B},  {\bf 53}
5149-5152 (1996).

\bibitem{morr}
Morr, D.K. and Pines, D. The resonance peak in cuprate superconductors.
{\em preprint}, cond-mat/9805107.




\bibitem{Hayden2} 
Hayden, S. {\it et al.} Absolute measurements of the high-frequency
magnetic dynamics in high-Tc superconductors. 
{\em Physica B}  {\bf 241-243} 765-772 (1998)

\bibitem{Fong}
Fong, H.F. {\it et al.}
Superconductivity-induced anomalies in the spin excitation spectra 
of underdoped $YBa_2 Cu_3 O_{6+x}$. {\em  Phys. Rev. Lett.} {\bf 78}
713-716 (1997)

\bibitem{Loram3} Loram, J.W. {\it et al.}
Superconducting and normal state energy gaps in
$Y_{0.8}Ca_{0.2}Ba_2Cu_3O_{7-\delta}$ from the electronic specific
heat. {\em Physica C} {\bf 282-287}  1405-1406 (1997).




\bibitem{Dai} Dai, P.  {\it et al.} Magnetic dynamics in undedoped
$YBa_2 Cu_3 O_{7-x} $: direct observation of a superconducting gap.
{\em  Phys. Rev. Lett.} {\bf 77} 5425-5428 (1996)



\end{thebibliography}
\end{document}